\documentclass[aps,prb,superscriptaddress,amsmath,amssymb,
floatfix,twocolumn]{revtex4-2}

\usepackage{amsmath,amssymb}
\usepackage{graphicx}

\usepackage{natbib}
\usepackage{hyperref}
\usepackage{siunitx}
\usepackage{booktabs}
\usepackage{xcolor}
\usepackage{subfigure}
\usepackage{epstopdf}
\usepackage{multirow}
\usepackage{ulem}

\newcommand{\spins}{\boldsymbol{S}}
\begin{document}
	
	\title{Determination of ground states of one-dimensional quantum systems using the cluster iTEBD method}
	
	\author{Tao Yang}
	\affiliation{National Laboratory of Solid State Microstructures and Department of Physics, Nanjing University, Nanjing 210093, China}
	\author{Rui Wang}
    \email{rwang89@nju.edu.cn}
	\affiliation{National Laboratory of Solid State Microstructures and Department of Physics, Nanjing University, Nanjing 210093, China}
	\affiliation{Collaborative Innovation Center for Advanced Microstructures, Nanjing University, Nanjing 210093, China}
    \affiliation{Hefei National Laboratory, Hefei 230088, China}
    \affiliation{Jiangsu Physical Science Research Center, Nanjing University, Nanjing 210093, China}
	\author{Z. Y. Xie}
    \email{qingtaoxie@ruc.edu.cn}
	\affiliation{School of Physics, Renmin University of China, Beijing 100872, China}
	\affiliation{Key Laboratory of Quantum State Construction and Manipulation (Ministry of Education),
		Renmin University of China, Beijing 100872, China}
	\author{Baigeng Wang}
	\affiliation{National Laboratory of Solid State Microstructures and Department of Physics, Nanjing University, Nanjing 210093, China}
	\affiliation{Collaborative Innovation Center for Advanced Microstructures, Nanjing University, Nanjing 210093, China}
    \affiliation{Jiangsu Physical Science Research Center, Nanjing University, Nanjing 210093, China}
		
	\date{\today}
	
	\begin{abstract}
		Within the framework of imaginary-time evolution for matrix product states, we introduce a cluster version of the infinite time-evolving block decimation algorithm for simulating quantum many-body systems, addressing the computational accuracy challenges in strongly correlated physics. By redefining the wave function ansatz to incorporate multiple physical degrees of freedom, we enhance the representation of entanglement, thereby improving the accuracy of the ground states. Utilizing the Trotter-Suzuki decomposition and optimized truncation schemes, our method maintains roughly the same computational complexity while capturing more quantum correlations. We apply this approach to three nontrivial cases: the gapless spin-1/2 Heisenberg chain, the spin-1 anisotropic XXZD chain with a higher-order Gaussian-type phase transition, and a spin-1/2 twisted triangular prism hosting a magnetic plateau phase. Improved accuracy in physical quantities, such as magnetization, ground state energy, and entanglement entropy, has been demonstrated. This method provides a scalable framework for studying complex quantum systems with high precision, making it suitable for situations where a pure increase in bond dimension alone cannot guarantee satisfactory results.
	\end{abstract}
	
	\maketitle
	
	\section{Introduction}
	The simulation of quantum many-body systems in strongly correlated physics poses significant computational challenges due to the exponential growth of the Hilbert space with system size. In the past decades, efficient algorithms with only algebraic complexity have been developed, such as density matrix renormalization group (DMRG) \cite{DMRG1992, DMRG2004} and time-evolving block decimation (TEBD) \cite{TEBD, iTEBD}. Though designed mainly for one-dimensional quantum systems, these methods have been successfully extended to more complex geometries, such as cylinders, torus shapes, and even two-dimensional systems \cite{Miles2012, JHC2008}. Particularly, extending the underlying concept of matrix product state (MPS) to higher dimension \cite{PEPS2004, OrusReview2019} and employing the translation symmetry, TEBD algorithms have shown great potential in the framework of imaginary-time evolution in dealing with infinite two-dimensional quantum systems, such as topological order \cite{JWMei2017, YJKao2022, RuiWang}, quantum spin liquid \cite{Kagome2017, KimNC2020}, superconductivity \cite{CorbozSC, SY2023}, and so on. 
		
	In the infinite TEBD (iTEBD) algorithm \cite{iTEBD}, the bond dimension ($\chi$) of the MPS representation of the ground state is an important hyperparameter that controls the number of variables, the upper-bound of the captured entanglement entropy, and thus the accuracy of the calculation. Therefore, in addition to designing more accurate update methods, such as canonical-form-based truncation and energy minimization variation, maintaining a larger bond dimension appears to be the only way to improve accuracy, especially when there is no suitable symmetry in the targeted systems. Therefore, it is valuable to develop a more efficient iTEBD algorithm whose performance can be improved from multiple aspects.
	
	In this work, we address this issue by introducing another hyperparameter, the cluster size ($n$) of the MPS ansatz, to the original iTEBD algorithm. In this modified version, the ground state is represented as a special MPS form, where each local tensor supports a local physical space corresponding to $n$ particles in an ordinary MPS representation. Meanwhile, the imaginary-time evolution operator $e^{-\tau H}$ is decomposed accordingly into local terms categorized as intra-cluster and inter-cluster parts, respectively. Then, the MPS is evolved using these local terms based on the singular value decomposition (SVD) truncation strategy, assisted by entanglement spectra, until convergence is reached at an accepted precision. The algorithm has a computational cost scaling as $O(\chi^2d^{n+2})+O(\chi^3d^{n+1})+O(\chi^3d^{6})$, where $d$ is the dimension of the Hilbert space spanned by each particle. This complexity is the same $O(\chi^3)$ in the large-$\chi$ limit as in the original iTEBD algorithm \cite{TEBD, iTEBD}, as long as $n$ is practically finite. 
	
	The validity of this cluster version of iTEBD, referred to as \emph{cluster iTEBD} henceforth in this work, is demonstrated in the spin-1/2 Heisenberg chain, where better ground state energy and magnetization density are obtained, compared to the original iTEBD with the same $\chi$. We then applied this modified algorithm to two other interesting models. One is the spin-1 XXZD chain, and the other is the spin-1/2 twisted triangular prism. We verify a third-order phase transition between a Haldane phase and a symmetry-breaking large-$D$ phase in the XXZD chain, and confirm a $1/3$-plateau phase surrounded by gapless phases in the prism model. In all these three cases, purely increasing the bond dimension $\chi$ cannot guarantee satisfying results, and this is the exact situation where the cluster size $n$ can play an important complementary part. 
		
	The rest of the paper is organized as follows. In Sec.~\ref{Sec: method}, we briefly review the original iTEBD algorithm and introduce the modified algorithm. In Sec.~\ref{Sec: models}, the results of the three models mentioned above are presented in detail. In Sec.~\ref{Sec: discuss}, we summarize and discuss some related topics.

	\section{Method}
	\label{Sec: method}
	In this section, we present the details of the algorithm proposed in this work. For this purpose, we first need to provide a brief review of the original iTEBD method, as proposed in Ref.~\cite{iTEBD}. 
	\subsection{Original iTEBD method}
	Employing the translation symmetry, the iTEBD method can be applied directly to the thermodynamic limit. Taking a one-dimensional translation-invariant system with only nearest neighbor interactions as an example, $H = \sum_{i}h_{i,i+1}$, for any tiny time step $\tau$, the imaginary-time evolution operator $e^{-\tau H}$ can be approximated by the Trotter-Suzuki decomposition \cite{Trotter1959}, 
	\begin{eqnarray}
		e^{-\tau H} \approx \prod_{i\in even} e^{-\tau h_{i,i+1}}\prod_{i\in odd} e^{-\tau h_{i,i+1}}.
	\label{Trotter}
	\end{eqnarray}
	The advantage of Eq.~(\ref{Trotter}) is that if the initial state $|\psi_0\rangle$ is represented as an MPS, described below, then $e^{-\tau H}|\psi_{0}\rangle$ can be evaluated efficiently with only local operations, and finally the ground state $|\psi_g\rangle$ can be obtained by applying $e^{-\tau H}$ repeatedly until some convergence is reached.
		
\begin{figure}
		\centering
		\includegraphics[width=\linewidth]{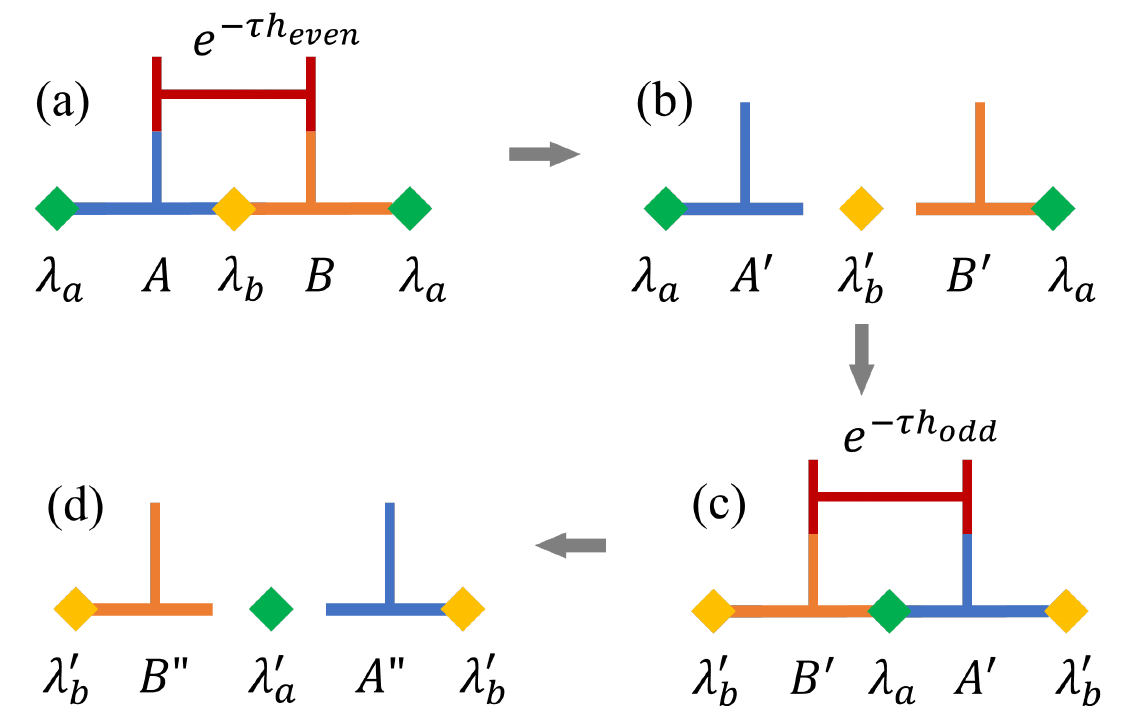}
		\caption{\label{fig: SU} Sketch of one iteration of the original iTEBD algorithm, and take the two-sublattice case as an example. For simplicity, $A$-$B$ bonds and $B$-$A$ bonds are denoted as even and odd, respectively. (a) Evolution of the MPS with $e^{-\tau h_{ij}}$ on even bond, where $\lambda_a$ is the effective bond vector defined on the odd bond. (b) Truncation based on the singular value decomposition of the local even cluster. (c)(d) Same as (a)(b), but evolution of the MPS with $e^{-\tau h_{ij}}$ on odd bonds.}
	\end{figure}

	Specifically, to compute $e^{-\tau H}|\psi_{0}\rangle$, the original iTEBD method represents $|\psi_{0}\rangle$ as an MPS \cite{MPS2007, Schollwoeck2011},
	\begin{eqnarray}
		|\psi_0\rangle = \textrm{Tr}\prod_{i\in odd}\lambda^{(i)}_a A_i[a_i]\lambda^{(i+1)}_bB_{i+1}[a_{i+1}]|...a_i...\rangle,
		\label{eq: MPS}
	\end{eqnarray}
	where we have assumed a two-sublattice structure, and $A_i$ and $B_i$ are local tensors supporting the physical space of the $i$-th particle. $\lambda^{(i)}_{a/b}$ is a positive vector defined on the $i$-th (odd or even) bond link, and its square mimics the entanglement spectra for the corresponding bipartition. Due to translation symmetry, the iTEBD method adopts an update strategy based on $\lambda$-assisted SVD truncation. The process is illustrated in Fig.~\ref{fig: SU}. Firstly, construct a local system by grouping a local operator $e^{-\tau h_{i,i+1}}$ for the even bond (e.g., $A$-$B$ bond), two local tensors $A$ and $B$ sharing the even bond, and the corresponding odd bond vector $\lambda_a$ and even bond vector $\lambda_b$, as shown in Fig.~\ref{fig: SU}(a). Then, bipartition the local system through SVD, from which the even bond vector $\lambda_b$ residing on the internal link of the cluster is updated as the obtained singular values. The local tensors are updated by the singular vectors appropriately, as shown in Fig.~\ref{fig: SU}(b). It is worth mentioning that in the above process, the odd bond vector $\lambda_a$ at the edge of the cluster remains unchanged, and it will be updated similarly when evolution for the odd bond (e.g., $B$-$A$ bond) is performed, as illustrated in Fig.~\ref{fig: SU}(c-d). After each SVD, to keep the bond dimension $\chi$ finite, the updated local tensors and entanglement spectra are truncated according to the singular values.
	
	The above procedure completes an iteration of the iTEBD algorithm. This iteration is performed repeatedly until some convergence of the entanglement spectra is reached, and the converged MPS can be regarded as an estimation of the actual ground state $|\psi_g\rangle$. This algorithm is very efficient, and from the perspective of DMRG, the bond vector $\lambda$ is usually regarded as \emph{effective environment} of the target system \cite{Schollwoeck2011} and plays a similar role to the effective mean field in the belief propagation algorithm \cite{BP1982}. This idea has been extended to the two-dimensional lattice under the name simple update \cite{JHC2008} and the mean-field version of the second renormalization group \cite{SRG}.

	\subsection{Cluster iTEBD method}
	
	In this work, we introduce another usable hyperparameter, namely cluster size $n$, to the iTEBD method, making it more flexible. Firstly, we divide the system into local clusters composed of $n$ particles, and a general local Hamiltonian can be recast as
	\begin{eqnarray}
		H = \sum_{\alpha}\sum_{i\in \alpha, j\in\alpha}h^{(\alpha)}_{i,j} + \sum_{\langle \alpha,\beta\rangle}\sum_{i\in \alpha, j\in\beta}h^{(\alpha,\beta)}_{i,j},
	\end{eqnarray}
	where $\langle \alpha,\beta\rangle$ means nearest neighbor cluster. The first part encompasses all interactions between particles residing within the same cluster, while the second part comprises interactions between neighboring clusters. Accordingly, the time evolution operator for a tiny Trotter step is approximated as a product of three parts,
	\begin{eqnarray}
		e^{-\tau H}\approx \prod_{\alpha} e^{-\tau h^{(\alpha)}}\prod_{\alpha\in odd} e^{-\tau h^{(\alpha,\alpha+1)}}\prod_{\alpha\in even} e^{-\tau h^{(\alpha,\alpha+1)}}, \nonumber \\
		\label{eq: Hdecom}
	\end{eqnarray}
	where $h^{(\alpha)}$ and $h^{(\alpha,\alpha+1)}$ contain all the intra-cluster interactions and the inter-cluster interactions, respectively. 
	
	In order to evaluate $e^{-\tau H}|\psi_0\rangle$, we similarly represent $|\psi_0\rangle$ as an MPS but with \emph{clustered} physical indices, i.e.,
	\begin{eqnarray}
		|\psi_0\rangle = \textrm{Tr}\prod_{\alpha\in odd}\lambda^{(\alpha)}_a A_\alpha[m_\alpha]\lambda^{(\alpha+1)}_bB_{\alpha+1}[m_{\alpha+1}]|...m_\alpha...\rangle \nonumber, \\
		\label{eq: CMPS}
	\end{eqnarray}
	which is the same as Eq.~(\ref{eq: MPS}) but replacing degrees of freedom $a_i$ for each particle by $m_\alpha$ for each cluster. Suppose $d$ is the dimension of the local space for each particle, and there are $n$ particles in each cluster, then the dimension supported by $m_{\alpha}$ is $d^n$. In this sense, the original MPS representation expressed in Eq.~(\ref{eq: MPS}) can be regarded as a special case, i.e., $n=1$.
	
\begin{figure}
		\centering
		\includegraphics[width=\linewidth]{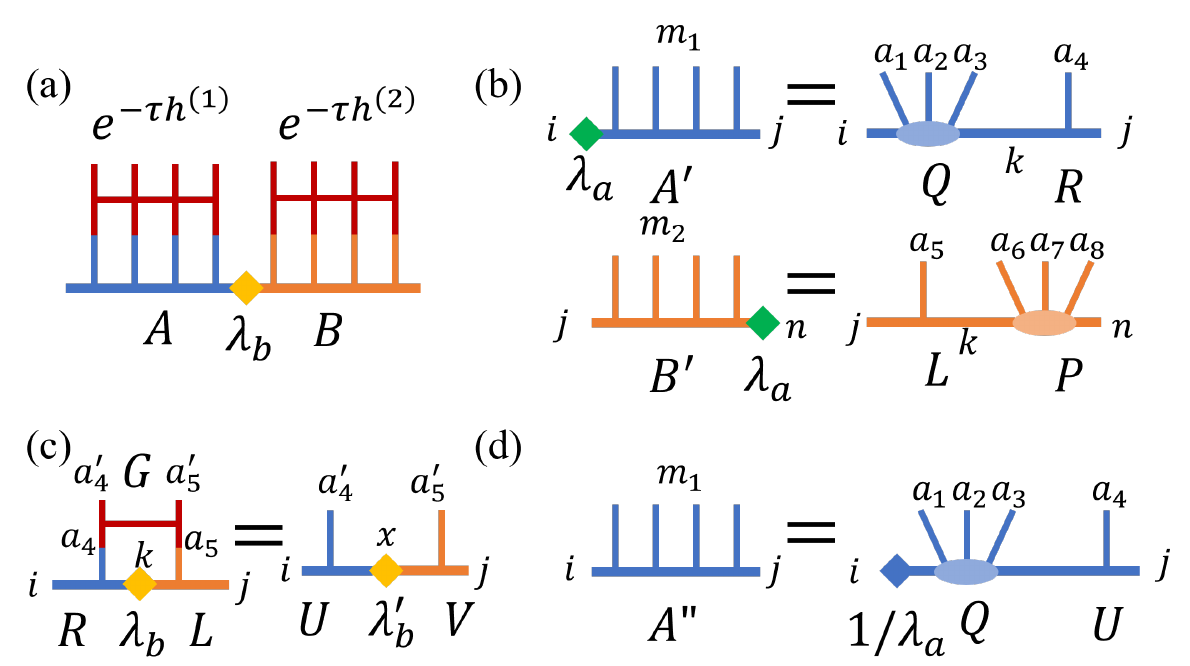}
		\caption{\label{fig: CSU} Sketch of the cluster version of the iTEBD algorithm, and take cluster size $n=4$ as an example. (a) Evolution of the MPS with $e^{-\tau h^{(1)}}$ and $e^{-\tau h^{(2)}}$, where $h^{(1/2)}$ denotes the Hamiltonian part which involves interactions among the four particles in the cluster labeled as 1 and 2. (b) Split the cluster tensors $A'$ and $B'$ into two subclusters through QR decomposition separately, as expressed in Eq.~(\ref{eq: QRD}). (c) Evolution of the MPS with $e^{-\tau h_{4,5}}$, where $h_{4,5}$ denotes the Hamiltonian part involving interaction between the two particles $a_4$ and $a_5$ residing at the neighboring edges. The local SVD is employed for truncation, as expressed in Eq.~(\ref{eq: SVD}). (d) The updated tensor $A''$ is recovered with the same structure with $A$. In this figure, the interactions are considered local for simplicity. }
	\end{figure}

	As long as Eqs.~(\ref{eq: Hdecom}) and (\ref{eq: CMPS}) are established, the iTEBD method can proceed with little extra effort compared to the original version. As illustrated in Fig.~\ref{fig: CSU}(a), the evolution with the first part in Eq.~(\ref{eq: Hdecom}) can be performed conveniently by absorbing $e^{-\tau h^{(\alpha)}}$ in the corresponding cluster directly. However, the evolution in the second part involves interactions between two neighboring clusters connected by an odd link, which requires some extra effort. Suppose $n=4$, and we need to evaluate $e^{-\tau h_{4,5}}$, where $a_4$ and $a_5$ are grouped into cluster indices $m_1$ and $m_2$, respectively. In order to save cost, as shown in Fig.~\ref{fig: CSU}(b), we construct a subsystem by absorbing $\lambda_{a}$ into local tensor $A' (B')$, and perform QR decomposition in the following
	\begin{eqnarray}
		A'_{i,j,m_1}\lambda_{a,i} &=& \sum_{k}Q_{ia_1a_2a_3,k}R_{k,a_4j}, \nonumber \\
		B'_{j,n,m_2}\lambda_{a,n} &=& 
		\sum_{k}P_{na_8a_7a_6,k}L_{k,a_5j},
		\label{eq: QRD}
	\end{eqnarray}  
	where $P$ and $Q$ are column-orthogonal matrices, and $L$ and $R$ are upper-triangular matrices. Then group $L$ and $R$ as a local system and perform SVD after the time evolution, i.e.,
	\begin{eqnarray}
		\sum_{k,a_4,a_5}R_{i,a_4k}\lambda_{b,k}L_{j,a_5k}G_{a_4a_5,a'_4a'_5} \approx \sum_{x\leq \chi}U_{ia'_4,x}\lambda'_{b,x}V_{ja'_5,x}, \nonumber \\
		\label{eq: SVD}
	\end{eqnarray}
	where $e^{-\tau h_{4,5}}$ is written explicitly as its matrix representation $G$, and the index $x$ is truncated according to the generated singular value $\lambda'_b$. As shown in Fig.~\ref{fig: CSU}(c-d), $\lambda'_b$ is regarded as the updated bond vector, and local tensors are updated by multiplying the inverse bond vector $1/{\lambda_a}$, the column-orthogonal matrices $Q$ ($P$), and $U$ ($V$) together. One can perform the evolution with the third part involving interactions between two neighboring clusters connected by an even link, similarly. 
	
	The above procedure completes an iteration of the cluster iTEBD algorithm. Besides the bond dimension $\chi$, it provides another hyperparameter, namely the cluster size $n$, which can be used to improve the accuracy. In some sense, the MPS ansatz used in Eq.~(\ref{eq: CMPS}) can be regarded as a one-dimensional version of the projected entangled simplex states \cite{PESS} proposed for higher-dimensional quantum systems, which are proved to be able to grasp more many-body entanglement than the conventional tensor network state in the context of imaginary-time evolution \cite{QianLi2022}. In the following, we will see that for a given $\chi$, within the framework of imaginary-time evolution, the MPS produced by the cluster iTEBD algorithm can capture more entanglement entropy than that obtained by the original iTEBD, and this could be the underlying reason why it can demonstrate better performance in our calculations.
			
	\section{Results}	
	\label{Sec: models}
	In this section, the cluster iTEBD method is tested in three nontrivial models. We first validate the method through the spin-1/2 Heisenberg chain. Compared to the original iTEBD method with the same bond dimension, we obtain a smaller magnetization density and a lower ground state energy. Then, we apply it to two interesting models, where we successfully verify a third-order topological quantum phase transition and confirm a $
1/3$-plateau phase, respectively. 
    
	\begin{figure}
		\centering
		\includegraphics[width=\linewidth]{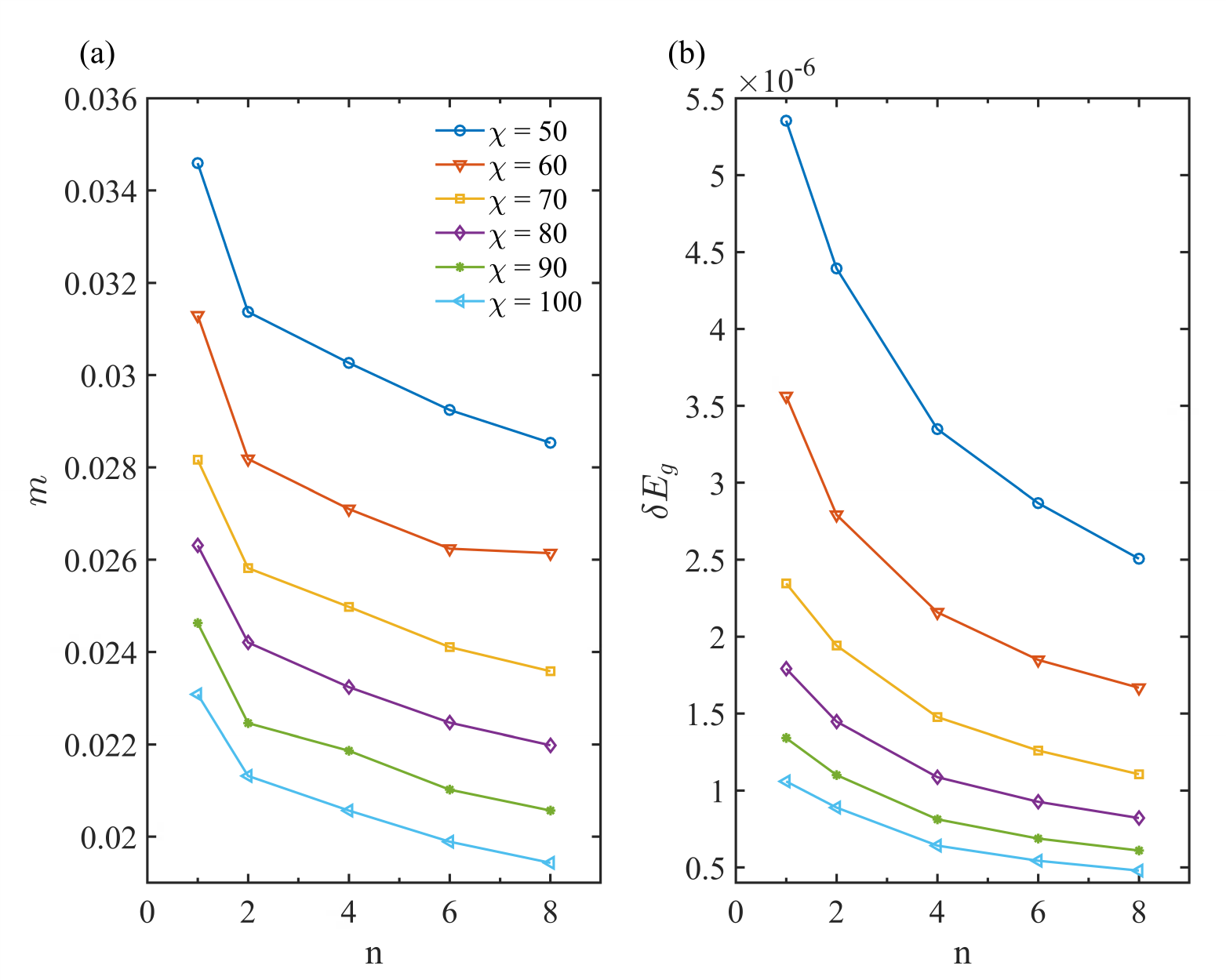}
		\caption{\label{fig: HeiPhys} Performance of the cluster iTEBD method for the spin-$1/2$ Heisenberg chain, with respect to $n$ (cluster size), for different $\chi$ (bond dimension of MPS). (a) Magnetization $m$. (b) Relative error $\delta E_g$ to the exact value of the ground state energy. The smallest Trotter step $\tau$ used is less than $10^{-3}$.}
	\end{figure}
	\subsection{Spin-1/2 Heisenberg Chain}
	\label{Sec: HeiChain}
	The spin-1/2 antiferromagnetic Heisenberg chain has been extensively studied long before. It is well known that the exact ground state energy is $E_{\textrm{exact}} = 1/4 - \ln{2} = -0.443147\cdots$ from Bethe ansatz \cite{MattisBook1988}, and that due to strong quantum fluctuations, the ground state is magnetically disordered from the Mermin-Wagner theorem \cite{MWtheorem}. However, since the system is gapless \cite{Haldane1983}, it is difficult to have an exact MPS representation of the ground state with finite bond dimension $\chi$. If SU(2) symmetry is not utilized explicitly, the MPS representation obtained from the iTEBD algorithm with a finite $\chi$ always possesses a weak but nonzero staggered magnetization. Similar phenomena have also been reported in two-dimensional systems \cite{PESS, LXC2022}. Therefore, this model provides an ideal platform to test whether the hyperparameter $n$ introduced in the cluster iTEBD method can improve the performance. 
    	\begin{figure}
		\centering
		\includegraphics[width=\linewidth]{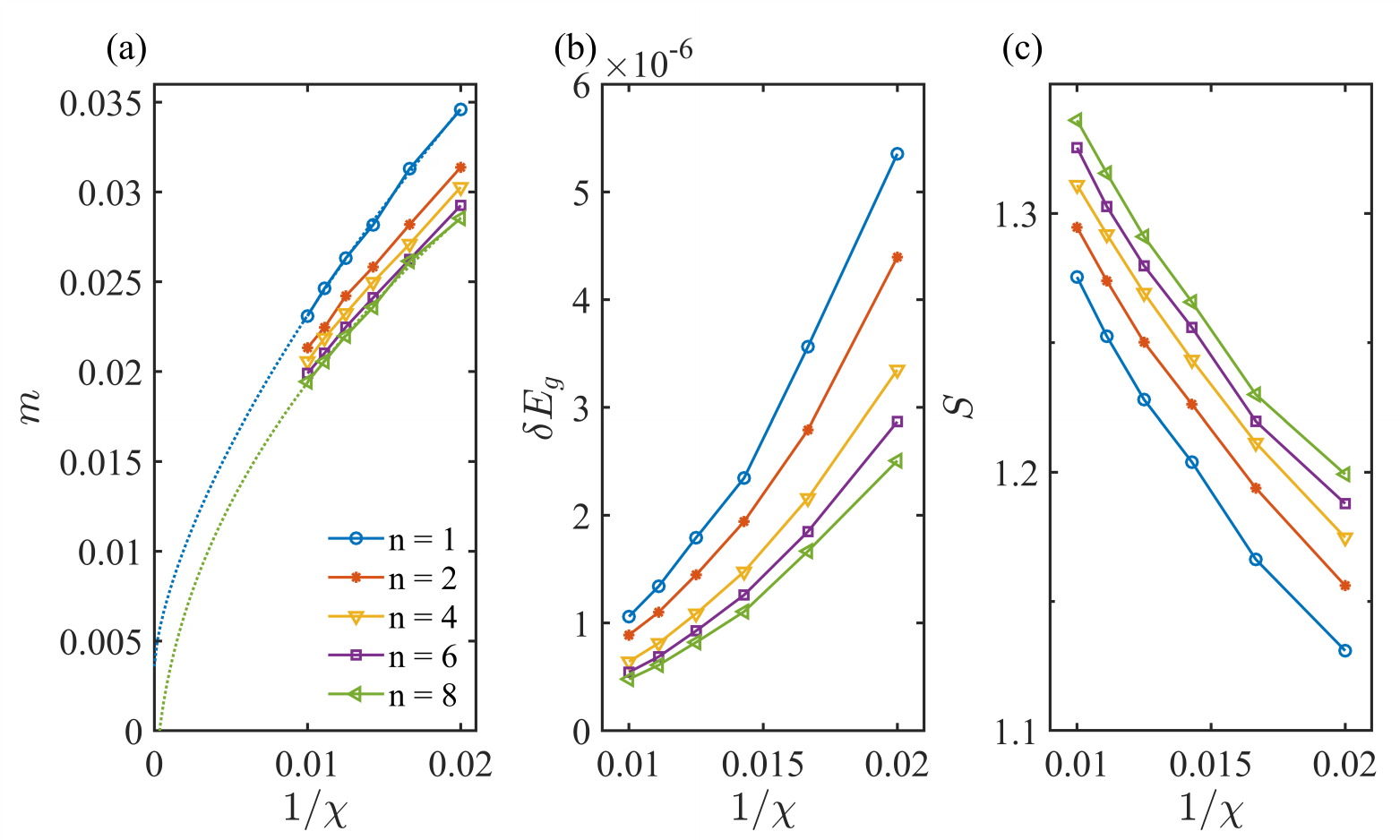}
		\caption{\label{fig: HeiScal} Performance of the cluster iTEBD method for the spin-$1/2$ Heisenberg chain, with respect to $\chi$ (bond dimension of MPS), for different cluster sizes $n$. (a) Magnetization $m$. The dashed lines indicate power-law fits to the data for $\chi \leq 100$. (b) Relative error $\delta E_g$ to the exact ground state energy. (c) Entanglement entropy $S$ obtained from the MPS representation for different parameters. Note: the case of $n=1$ corresponds to the original iTEBD.}
	\end{figure}
	The results are displayed in Fig.~\ref{fig: HeiPhys}. The smallest $\tau$ we used is less than $10^{-3}$ in our calculations. It shows that for any given cluster size $n$, the magnetization $m$ is indeed nonzero but decreases as the bond dimension $\chi$ becomes larger. Things are similar for the ground state energy $E_g$, i.e., for any given $n$, the relative error $\delta E_g\equiv\frac{E_{g}-E(\chi,n)}{E_g}$ decreases as $\chi$ becomes larger. More importantly, for a given $\chi$, the accuracy of the magnetization and the energy is systematically improved when $n$ is increased. This means that the cluster size $n$ indeed provides another usable control parameter to improve the performance of the iTEBD algorithm. This should be particularly useful when $\chi$ cannot be easily extended, e.g., in gapless systems where desirable $\chi$ might be quite large due to the area law of entanglement entropy \cite{AreaLaw}, or two-dimensional quantum systems where large $\chi$ is difficult to handle \cite{NTN2017}.
	
	Furthermore, in Fig.~\ref{fig: HeiScal}(a-b), we extrapolate the data of $m$ and $\delta E_g$ to the large-$\chi$ limit for any fixed $n$. It shows that, up to $\chi = 100$ and $n=8$, the cluster iTEBD algorithm can obtain a magnetization smaller than $0.02$ and energy accuracy about $5\times10^{-7}$, and the extrapolation for $n=8$ can give a much more accurate estimation of these quantities than that for $n=1$, i.e., the original iTEBD case. See the dashed lines in Fig.~\ref{fig: HeiScal}(a). This can be understood within the framework of imaginary-time evolution, since the cluster algorithm with the same bond dimension $\chi$ can produce an MPS that captures more quantum entanglement than the original one. The comparison of the captured entropy as a function of $n$ is demonstrated in Fig.~\ref{fig: HeiScal}(c). In practical calculations, if necessary, one can further extrapolate the obtained data $E(\chi=\infty, n)$ to the large-$n$ limit to obtain more accurate results.
		
	\subsection{Spin-1 Anisotropic XXZD Chain}
	\label{Sec: XXZD}
\begin{figure}
		\centering
		\includegraphics[width=0.65\linewidth]{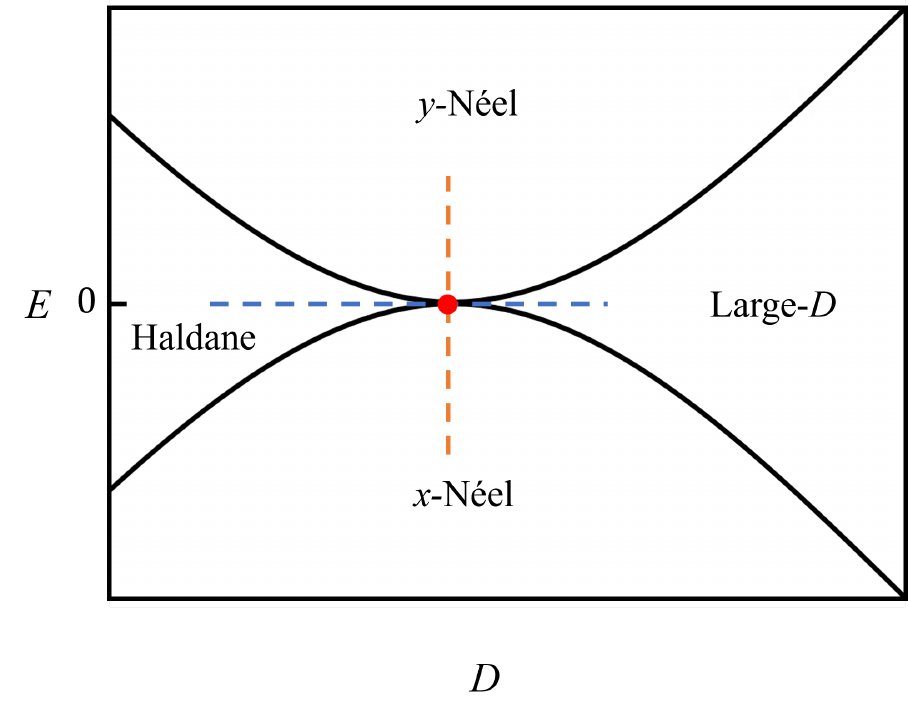}
		\caption{Cartoon illustration of phase diagram of the spin-1 XXZD chain with respect to uniaxial anisotropy $D$ and rhombic anisotropy $E$. The blue dotted line denotes a possible Gaussian-type third-order phase transition from the Haldane phase to a large-$D$ phase, while the orange dotted line denotes a second-order phase transition between two different N\'{e}el phases.}
		\label{fig: XXZD-PD}
	\end{figure}

	We now consider the spin-1 anisotropic XXZD chain. The model has XXZ-type interactions between spin-1 operators   $\spins_i$  defined on site $i$. Moreover,  the uniaxial and rhombic single-ion anisotropies are controlled by the strengths $D$ and $E$, respectively. The Hamiltonian is given by, 
	\begin{eqnarray}
		H ~=&& \sum_{\langle i,j \rangle} \left[S_i^x S_j^x+S_i^y S_j^y+J_zS_i^z S_j^z\right] \nonumber \\&+& D \sum_i (S_i^z)^2 + E \sum_i \left[ (S_i^x)^2 - (S_i^y)^2 \right],
	\end{eqnarray}
	When $E=0$, from theoretical analysis and numerical evidence \cite{YCT2008, LY2011, YJKao2017, JFY2023}, it is known that the topological phase transition from the Haldane phase to a so-called large-$D$ symmetry-breaking phase can be higher-order Gaussian type, e.g., third-order for $J_z = 1$ at $D_c\approx 0.9687$ and fifth-order for $J_z = 0.5$ at $D_c\approx 0.6197$ \cite{JFY2023}. Meanwhile, if we fix $D=D_c$ and vary $E$, there is an ordinary second-order phase transition between the two N\'{e}el phases breaking $Z_2$ symmetry, e.g., at $E=0$ when $J_z=1$. A sketch of the phase diagram for $J_z=1$ is illustrated in Fig.~\ref{fig: XXZD-PD}. This model has drawn attention recently, partly due to its relevance to a recent ultra-cold atom realization \cite{Chung2021} and inelastic neutron scattering experiments on real materials \cite{DM2022}.
	
	In this work, we set $J_z=1$ for simplicity, and focus on the two phase transitions, namely the third-order topological phase transition and the second-order symmetry-breaking one, as indicated by blue and orange dashed lines in Fig.~\ref{fig: XXZD-PD}, respectively. The third-order transition is not easily accessible. The previous calculations utilize complex quantities, such as fidelity \cite{YCT2008}, level crossing point \cite{YJKao2017}, or cross derivatives \cite{JFY2023}, to explore the signal, as the conventional second-order derivatives of the energy cannot yield a clear peak \cite{LY2011}. In this work, we expect a better identification of peaks in the usual second-order derivatives by increasing the hyperparameter $n$.
	
\begin{figure}
		\centering
		\includegraphics[width=\linewidth]{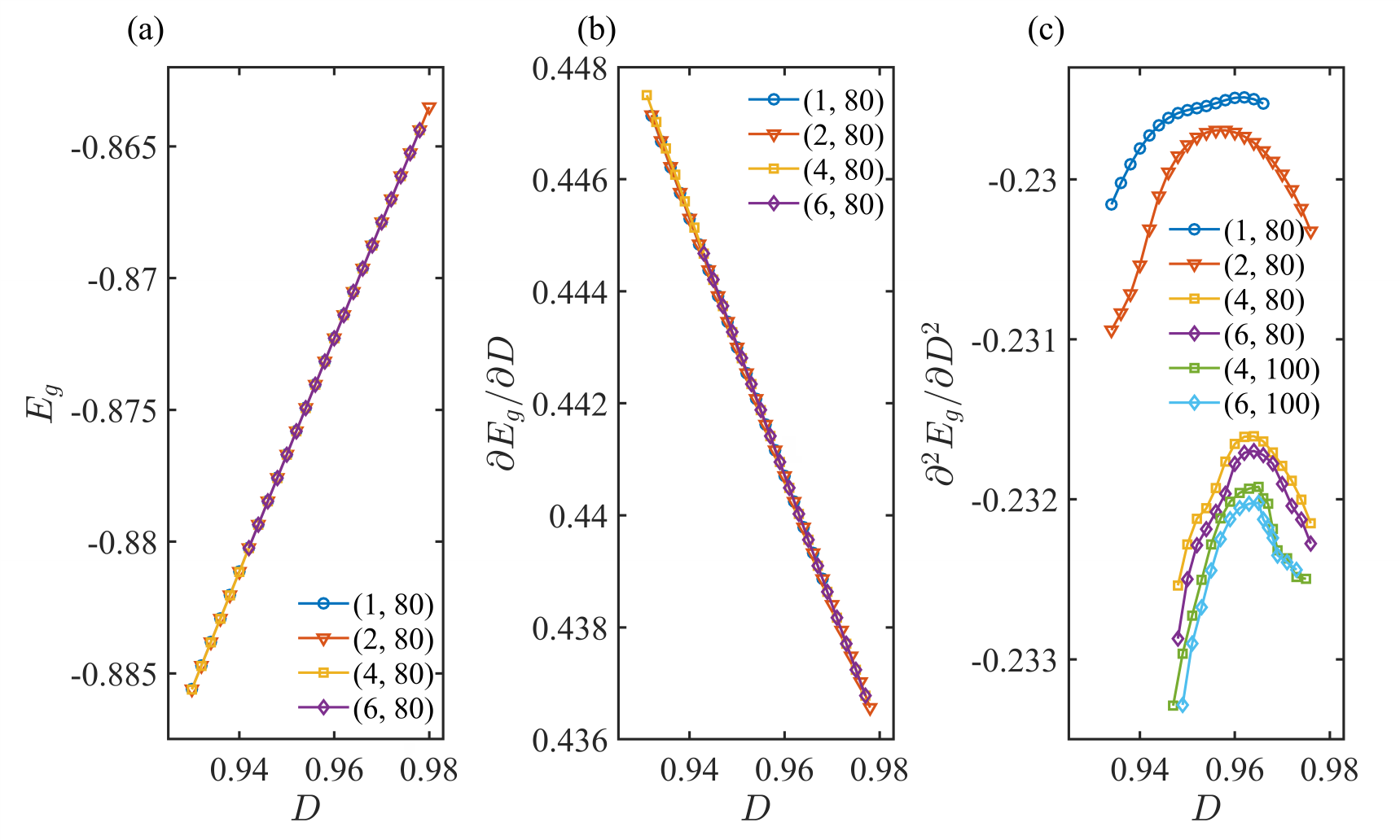}
		\caption{The behavior of (a) ground state energy $E_g$, (b) its derivative $\frac{\partial E_g}{\partial D}$, and (c) its second-order derivative $\frac{\partial^2E_g}{\partial D^2}$, with respect to $D$ along the blue dotted line ($E=0$) in Fig.~\ref{fig: XXZD-PD}, for the XXZD chain. The two parameters in the brackets of the legends represent ($n,\chi$) used in the calculations.}
		\label{fig: XXZD-3rd}
	\end{figure}

\begin{figure}
		\centering
		\includegraphics[width=\linewidth]{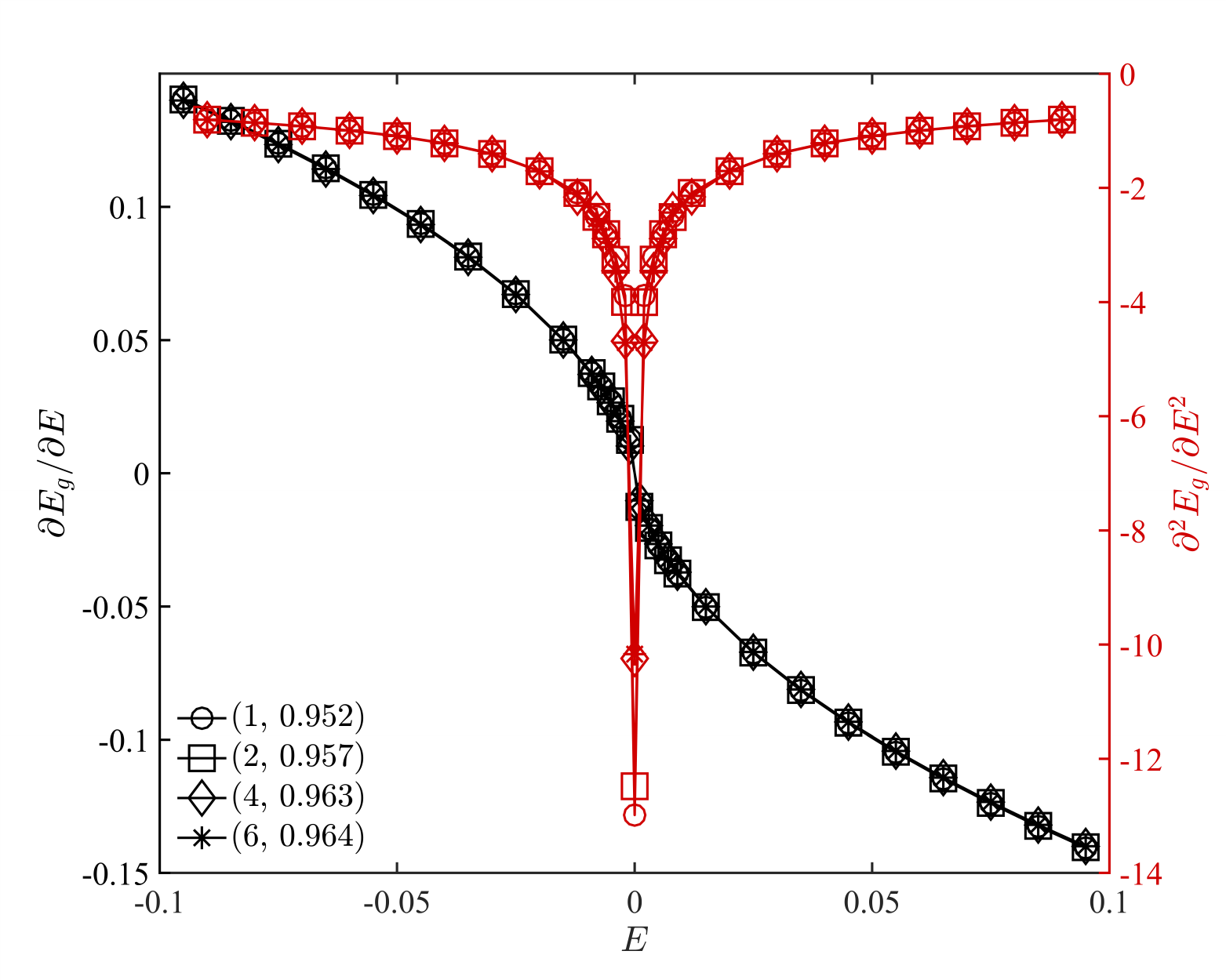}
		\caption{The behavior of the (a) derivative $\frac{\partial E_g}{\partial E}$ and (b) second-order derivative $\frac{\partial^2 E_g}{\partial E^2}$ of the ground state energy, with respect to rhombic anisotropy $E$ along the orange dotted line (with $D$ fixed as critical value $D_c$ obtained for a given $n$) in Fig.~\ref{fig: XXZD-PD}, for the XXZD chain. The two parameters in the brackets of the legends represent ($n, D_c$) used in the calculations. }
		\label{fig: XXZD-2nd} 
	\end{figure}

	The results are summarized in Fig.~\ref{fig: XXZD-3rd} and Fig.~\ref{fig: XXZD-2nd}. In Fig.~\ref{fig: XXZD-3rd}, setting $E=0$, we plot the ground state energy $E_g$ as a function of $D$ as well as its derivative $\frac{\partial E_g}{\partial D}$ and second derivative $\frac{\partial^2 E_g}{\partial D^2}$. It shows that, indeed, there are no obvious singularities in the curves; instead, the $\frac{\partial^2 E_g}{\partial D^2}$ develops a broad peak that becomes more and more significant as either $n$ or $\chi$ increases. The location of the peak for $(n=6,\chi=100)$ is about $D\approx 0.964$, providing a consistent estimation of the third-order phase transition with previous studies \cite{JFY2023}. On the contrary, as shown in Fig.~\ref{fig: XXZD-2nd}, by fixing $D=D_c$ obtained for each ($n,\chi$) setup and plotting the derivatives with respect to the rhombic anisotropy $E$, we can see clear singularities of $\frac{\partial^2 E_g}{\partial E^2}$ at $E=0$ for a series of $(n,\chi)$ combinations. This is consistent with the orange dashed line in Fig.~\ref{fig: XXZD-PD} indicating the second-order phase transition.
	
	We can see from this calculation that for the ordinary continuous phase transition with symmetry breaking which can be determined by the original iTEBD algorithm, the cluster iTEBD can provide more accurate critical information, such as the critical value $D_c$ as shown in Fig.~\ref{fig: XXZD-3rd}(c) and indicated in the legends of Fig.~\ref{fig: XXZD-2nd}. While for the Gaussian-type transition, where there is no singularity in the second-order derivatives, the iTEBD algorithm with $n=1$ cannot determine the phase transition easily, and the cluster iTEBD with larger $n$ can display peaks more clearly. 
	
	\subsection{Spin-1/2 Twisted Triangular Prism}
	\label{Sec: Tube}
\begin{figure}
		\centering
		\includegraphics[width=\linewidth]{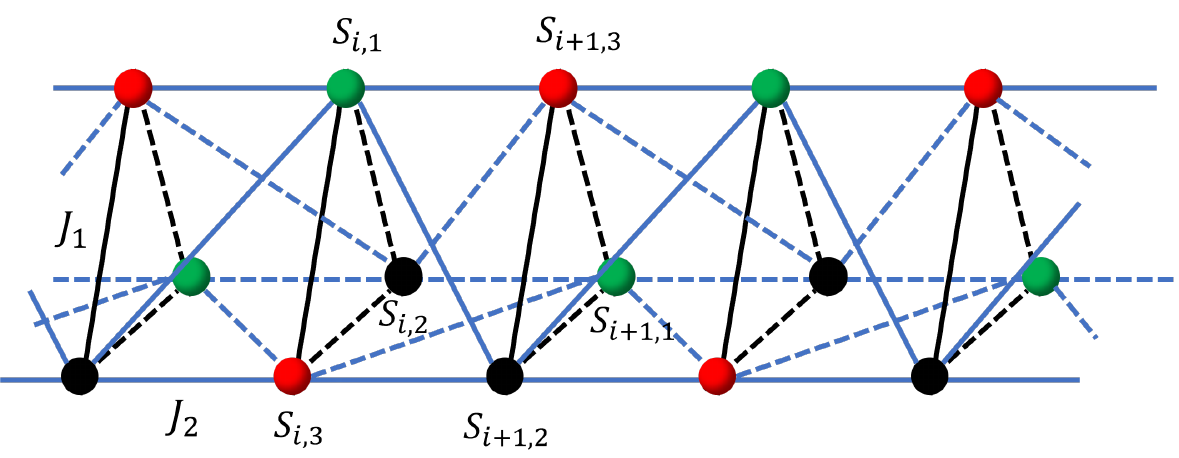}
		\caption{Sketch of a spin-$1/2$ twisted triangular prism composed of three spin chains with inter-chain interactions. Every spin has been assigned a color. Besides interacting with the spins in the same triangle, each spin interacts with those in the nearest neighbor triangles, but with different colors. The intra-triangle and inter-triangle interaction strengths are denoted as $J_1$ (black) and $J_2$ (blue), respectively. The solid and dashed lines are used solely to enhance the stereoscopic effect, and as long as the colors are the same, they essentially represent the same interaction.}
		\label{fig: Tube}
	\end{figure}

	The third model we studied is a spin-1/2 twisted triangular prism, which can be effectively depicted in Fig.~\ref{fig: Tube}. The triangular prism is composed of three spin chains (horizontal lines), and the three spins on the same cross-section construct a triangle. The spins on the same triangle are assigned different colors, and by \emph{twist}, we mean that each spin interacts with only those of different colors in the same or nearest neighbor triangles. The Hamiltonian can be effectively expressed as	
	\begin{eqnarray}
		H = J_1 \sum_{i;j\neq k} \spins_{i,j} \cdot \spins_{i,k} 
		+J_2 \sum_{i;j\neq k}\spins_{i,j}\cdot\spins_{i+1,k} - h\sum_{i,j}S_{i,j}^z, \nonumber \\
	\end{eqnarray}
	where $\spins_{i,j}$ is the spin-1/2 operator defined on the $i$-th triangular, and $j = 1, 2, 3$ denotes its color index. $J_1$ means interaction strength in the same triangle (denoted by black in Fig.~\ref{fig: Tube}), $J_2$ denotes interaction strength between nearest neighbor triangles (denoted by blue in Fig.~\ref{fig: Tube}), and $h$ represents a uniform external magnetic field in the $z$-direction. Obviously, if the system is considered a one-chain system, there must exist interactions between at least third-nearest neighbors, which is not easy for the conventional iTEBD to handle.
	
	In this work, we set the energy scale as $J_1 = 1$ for simplicity. Due to the strong frustration, the model hosts a rich phase diagram in the range $0 < J_2 \lesssim 1.22$ \cite{Mila2006, JPSJ2015}. Specifically, when turning on $h$ within this parameter range, as determined by DMRG and exact diagonalization on finite systems, the model was found to host a $1/3$-plateau state surrounded by gapless magnetically ordered phases \cite{Mila2006}. The model is believed to be relevant to some real compounds, such as $[(\textrm{CuCl}_2\textrm{tachH})_3\textrm{Cl}]\textrm{Cl}_2$ \cite{Cronin2004}, and it is argued to apply to other frustrated spin tubes as well. 
	
	Using the cluster iTEBD algorithm, the MPS ansatz in Eq.~(\ref{eq: CMPS}) suitable for this system can be conveniently represented by grouping the spin degrees of freedom in each triangle as a single \emph{composite} particle, which spans a local Hilbert space with dimension $2^3=8$. Then, the intra-triangle and inter-triangle interactions are reduced to just on-site and nearest-neighbor ones, respectively. In this work, we use the minimal $n=3$ for this model in all the calculations.
	
\begin{figure}
		\centering
		\includegraphics[width=\linewidth]{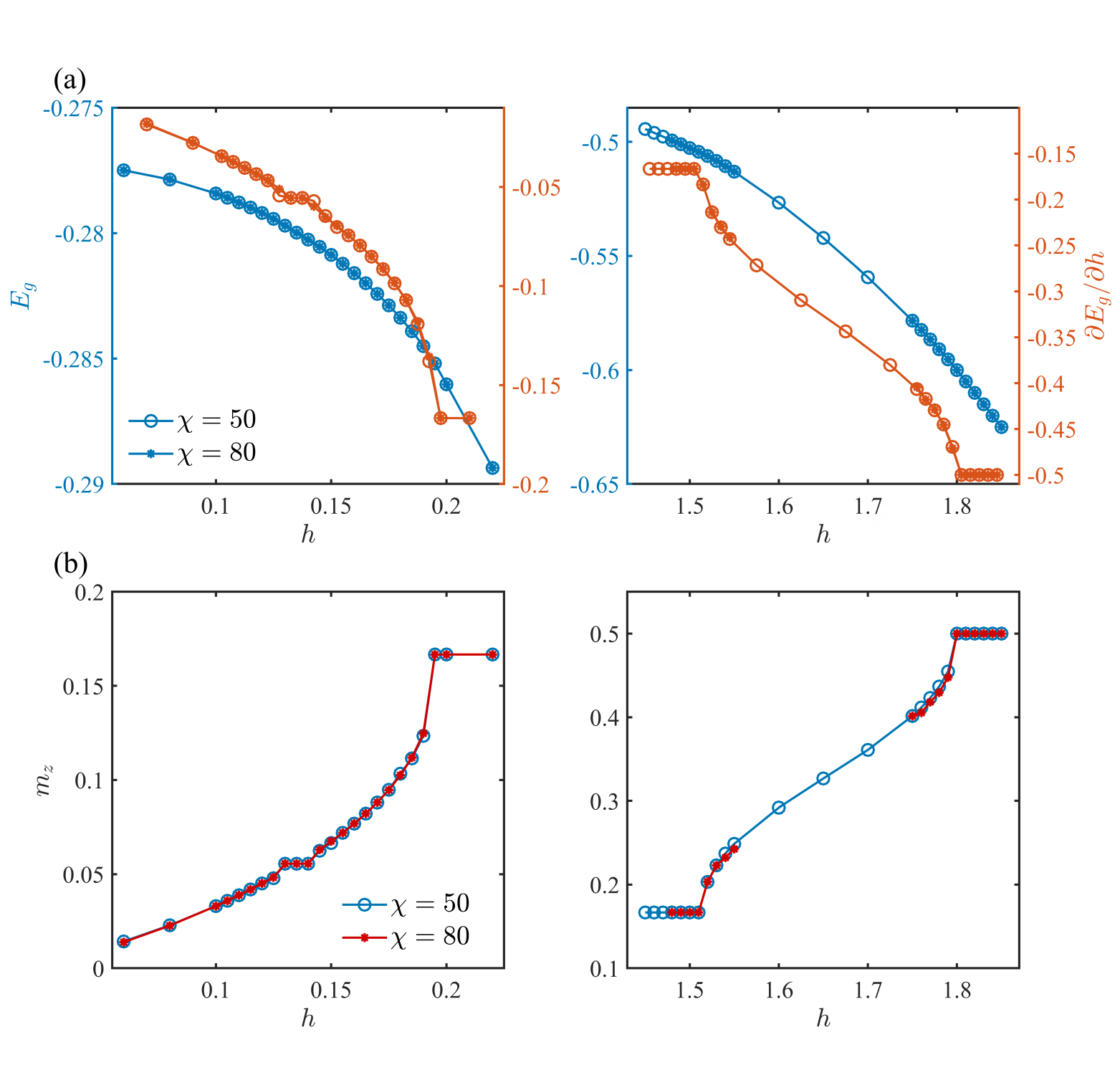}
		\caption{The behavior of (a) ground state energy $E_g$ and its derivative $\frac{\partial E_g}{\partial h}$ with respective to external field $h$, and (b) the magnetization $m_z$, for the spin prism with $J_2 = 0.1$. }
		\label{fig: Tube-Start}
	\end{figure}

\begin{figure}
		\centering
		\includegraphics[width=\linewidth]{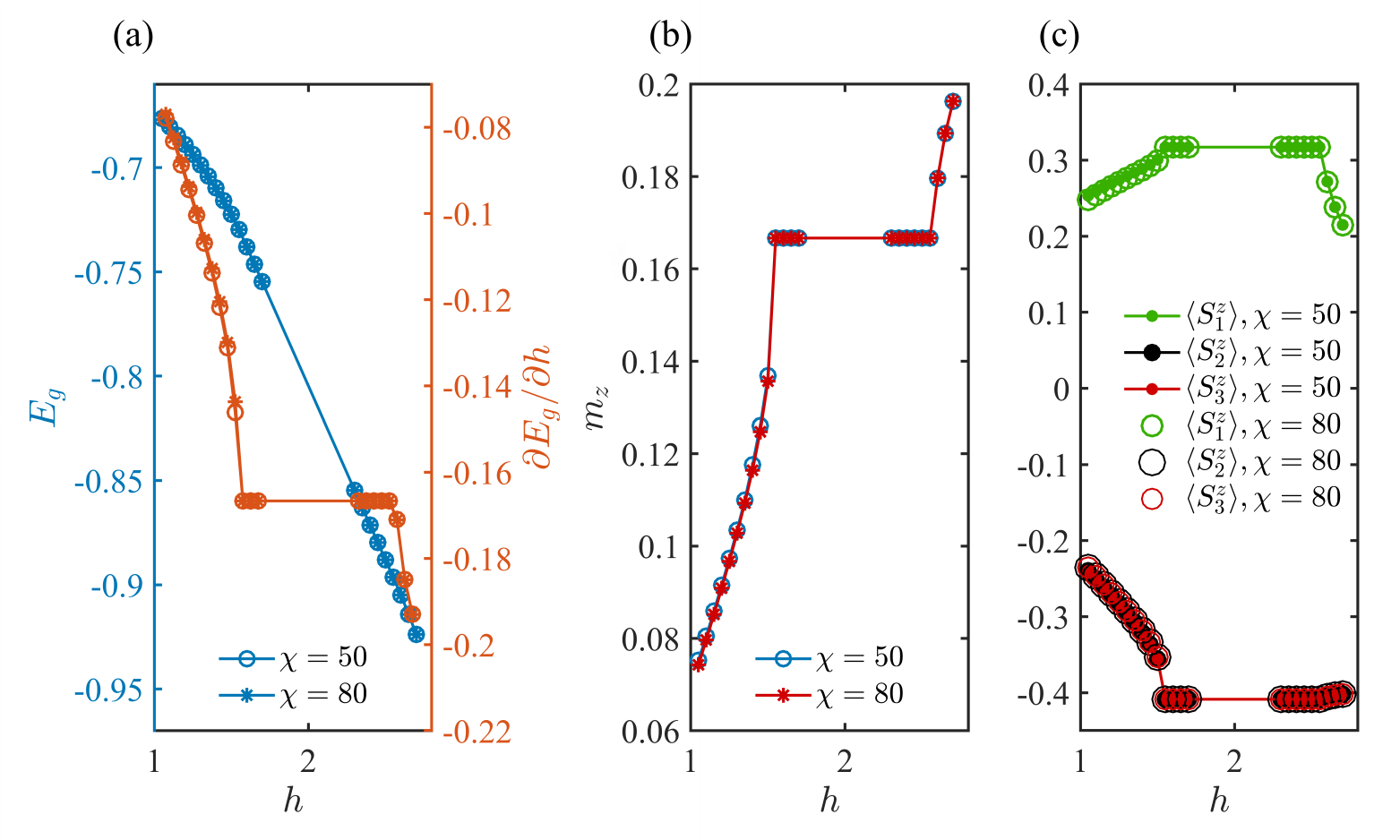}
		\caption{The $1/3$ magnetic plateau phase in the spin prism model with $J_2 = 1.2$. (a) Ground state energy $E_g$ and its derivative $\frac{\partial E_g}{\partial h}$. (b) Magnetization $m_z$. (c) Magnetic configuration in each triangle.}
		\label{fig: Tube-End}
	\end{figure}

	We focus on two specific parameters, namely $J_2 = 0.1$ and $J_2 = 1.2$, which are close to the two edges of the interesting range mentioned above. The results are illustrated in Fig.~\ref{fig: Tube-Start} and Fig.~\ref{fig: Tube-End}. For $J_2 = 0.1$, we calculate the ground state energy $E_g$, its derivative $\frac{\partial E_g}{\partial h}$, as well as magnetization $m_z$, as functions of magnetic field $h$. It shows that the derivative is a constant over a wide range about $0.195\lesssim h \lesssim 1.51$, and this is consistent with the fact that the magnetization displays a $\frac{m}{1/2}\approx \frac{1}{3}$ plateau at the same range. The plateau in $\frac{\partial E_g}{\partial h}$ at $h\gtrsim 1.8$ corresponds to the fully polarized state residing at the magnetic plateau with $m_z=0.5$. For $J_2 = 1.2$, the derivative $\frac{\partial E_g}{\partial h}$ and $m_z$ also show plateau behavior but at the range of $1.55\lesssim h\lesssim 2.55$. Furthermore, to reveal the nature of the $1/3$-plateau state, in Fig.~\ref{fig: Tube-End}(c), we plot the magnetization density for each spin separately in the same triangle. We find that the spins denoted by black and red ($\spins_{i, 2}$ and $\spins_{i, 3}$) tend to be parallel to each other, while spins colored green ($\spins_{i, 1}$) are always anti-parallel to them. This should correspond to the so-called UUD state found in Ref.~\cite{JPSJ2015}, and we do not find a plateau state with uniform magnetization claimed there.
	
\begin{figure}
		\centering
		\includegraphics[width=\linewidth]{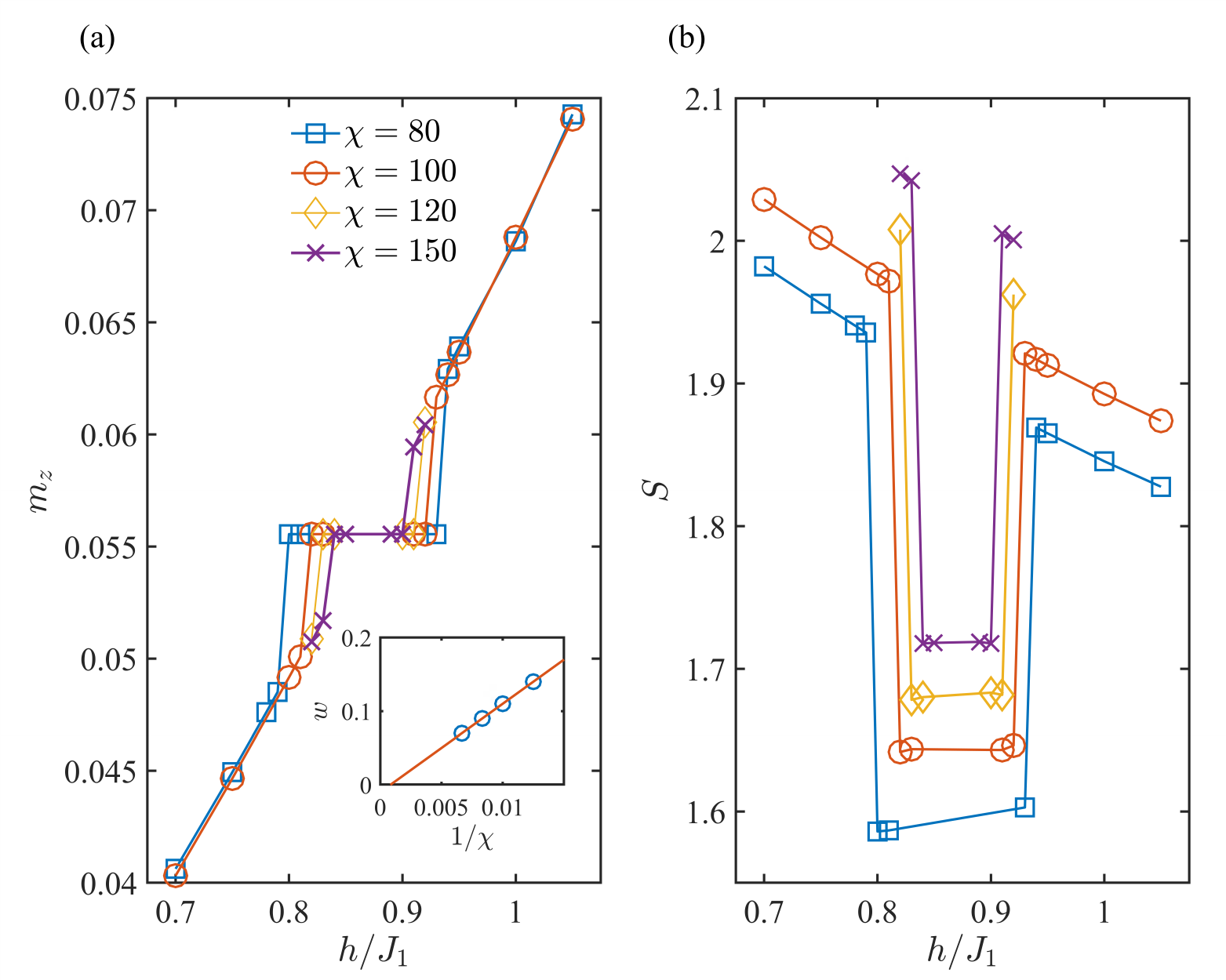}
		\caption{\label{fig: Plateau-EE} The metastable $1/9$-plateau in the spin tube with $J_2 = 1.2$. (a) Ground state energy $E_g$ and magnetization $m_z$. (b) Entanglement entropy $S$ obtained from the inter-cluster entanglement spectra in the MPS representations.}
	\end{figure}

	In fact, in Fig.~\ref{fig: Tube-Start}, besides the $1/3$-plateau and saturated plateau, there also seems to exist a $1/9$-plateau phase at the range of $0.13\lesssim h\lesssim 0.14$. Similar things happen for $J_2=1.2$ at about $0.8\lesssim h\lesssim 0.9$. To verify this possibility, we extend our calculation to $\chi = 150$, and the result is summarized in Fig.~\ref{fig: Plateau-EE}(a) for $J_2 = 1.2$. It shows clearly that as $\chi$ becomes larger, the width $w$ of the plateau becomes narrower and narrower. This is confirmed further by the entanglement entropy extracted from the MPS representation of the ground state, as shown in Fig.~\ref{fig: Plateau-EE}(b): the entanglement entropy is strongly suppressed in the plateau phase, and the range of the suppressed region becomes smaller as $\chi$ becomes larger. What's more, in the inset of Fig.~\ref{fig: Plateau-EE}(a), we find that the data of $w$ can be well fitted by linear functions of $1/\chi$, and the extrapolation to the large-$\chi$ limit gives $w\simeq 0$. Therefore, the obtained $1/9$-plateau is probably a metastable state arising from the finite-$\chi$ effect and should be absent in the true ground state. 
		
	\section{Summary and Discussion}
	\label{Sec: discuss}
	
	In this study, we have proposed a cluster version of the iTEBD algorithm to accurately determine the ground state of a one-dimensional quantum lattice system. Similar to the original one that has been widely used, the cluster algorithm involves designing a clustered MPS ansatz, decomposing the imaginary-time evolution operator as intra- and inter-cluster parts accordingly, and updating the MPS by entanglement-spectra-assisted SVD truncation. The cluster size $n$, as well as the bond dimension $\chi$, provides useful control parameters for improving the performance. Its validity has been demonstrated in the spin-1/2 antiferromagnetic Heisenberg chain, where the cluster iTEBD can provide a lower ground state energy and smaller magnetization compared with the original iTEBD with the same bond dimension. After that, we first apply it to the spin-1 anisotropic XXZD chain. Through the second derivatives $\frac{\partial^2 E_g}{\partial D^2}$, we verify a third-order Gaussian-type phase transition between the Haldane phase and the large-$D$ phase, as well as a second-order phase transition between two N\'{e}el phases. Then, we apply it to the spin-1/2 twisted triangular prism, and identify a $1/3$-plateau phase and metastable $1/9$-plateau phase through magnetization and entanglement entropy calculations.
	
	Given cluster size $n$, bond dimension $\chi$, and degrees of freedom $d$ of a single particle, the cluster iTEBD algorithm has computational cost scaling as $O\left(\chi^2d^{2n}\right)+O\left(\chi^3d^{n+1}\right)+O\left(\chi^3d^{6}\right)$, if neither symmetry nor partial SVD has been considered. The first part comes from intra-cluster evolution and can be reduced to $O\left(\chi^2d^{n+2}\right)$ by further decomposing the $e^{-\tau h^{(\alpha)}}$ bond by bond separately, instead of regarding it as a whole. The remaining two parts come from inter-cluster evolution involving Eq.~(\ref{eq: QRD}) and Eq.~(\ref{eq: SVD}), respectively. When $n$ is finite, in the large-$\chi$ limit, the cluster algorithm has the same order of $\chi$ as the original iTEBD, namely $\chi^3$. 
	
	Capturing the entanglement entropy more easily and being able to systematically improve performance by enlarging $n$, the cluster iTEBD is advantageous in at least two situations. One is the situation where purely increasing $\chi$ is difficult to achieve satisfactory results, such as fast convergence of physical quantities in gapless systems like the spin-1/2 Heisenberg chain discussed in Sec.~\ref{Sec: HeiChain}, and pronounced peaks in derivatives like the spin-1 XXZD chain studied in Sec.~\ref{Sec: XXZD}. The other is the system hosting interactions with longer ranges than nearest neighbors, e.g., the spin-1/2 twisted triangular prism described in Sec.~\ref{Sec: Tube}. When dealing with longer-range interactions, such as $h_{2,6}$ for $n=4$ ansatz depicted in Fig.~\ref{fig: CSU}(a), to perform the time evolution of $e^{-\tau h_{26}}$, Eq.~(\ref{eq: QRD}) can be modified similarly by separating $a_2$ and $a_6$ from other indices, i.e.,
        \begin{eqnarray}
		A'_{i,j,m_1}\lambda_{a,i} &=& \sum_{k}Q_{ia_1a_3a_4,k}R_{k,a_2j}, \nonumber \\
		B'_{j,n,m_2}\lambda_{a,n} &=& 
		\sum_{k}P_{na_8a_7a_5,k}L_{k,a_6j}.
		\label{eq: QRD1}
	\end{eqnarray}  
We note that the form of Eq.~(\ref{eq: SVD}) is unchanged, but the indices $a_4$ and $a_5$ are now replaced by $a_2$ and $a_6$, respectively. Then, the subsequent procedures remain the same as before. In fact, it is easy to show that as long as the interaction range is no larger than $n$, the cluster iTEBD algorithm with cluster size $n$ involves only the nearest neighbor interactions.

	The cluster iTEBD algorithm shares some resemblance with the projected entanglement simplex state method \cite{PESS} and the so-called regularized time evolution scheme \cite{LXC2022}, in the sense that in these work, multiple degrees of freedom are also grouped together in either wave function ansatz or time evolution operator. It is also worth mentioning that, in fact, it has little to do with previous studies of \emph{cluster update}, which updates a two-dimensional state in the imaginary-time evolution by regarding a small cluster nearby as the effective environment of a target system \cite{CU2011}. In general, there are no grouped degrees of freedom in the cluster update algorithm.
	
	In this work, the maximal kept bond dimension $\chi$ is no more than 150, because it is already sufficient for our specific purpose. However, it is worth mentioning that since the cluster iTEBD method can produce an MPS with more entanglement than the original one (though the upper bound $\log\chi$ remains unchanged), as shown in Fig.~\ref{fig: HeiScal}(c), it is necessary to use larger $\chi$ to obtain the same truncation error as the original iTEBD. In other words, the cluster iTEBD algorithm provides a better estimation of how large $\chi$ should be used for a given system, and it makes no sense to use a large $n$ but with a small $\chi$. In practical calculations, one should make a balance between $n$ (entanglement) and $\chi$ (truncation error), which is similar to the case of two-dimensional quantum systems \cite{NTN2017}. In any case, keeping a sufficient $\chi$ is always necessary, as alerted in Fig.~\ref{fig: Plateau-EE} for the $1/9$ magnetic plateau in the prism model.
	
	\section*{Acknowledgments}
    We are grateful to Rui-Zhen Huang, Hongyu Chen, and Rong Yu for their helpful discussions. This work was supported by the National R\&D Program of China (Grants No. 2023YFA1406500, 2024YFA1408604, 2022YFA1403601, 2024YFA1410500), the National Natural Science Foundation of China (Grants No. 12274458, No. 12322402, No. 12274206), the Innovation Program for Quantum Science and Technology (Grant No. 2021ZD0302800), the Natural Science Foundation of Jiangsu Province (No. BK20233001), and the Xiaomi foundation.


\end{document}